\documentclass[fleqn,usenatbib]{mnras}

\usepackage[english]{babel}
\usepackage{lipsum}
\usepackage{placeins}
\usepackage{newtxtext,newtxmath}
\usepackage[T1]{fontenc}
\usepackage{ae,aecompl}
\usepackage{graphicx}	
\usepackage{amsmath}	
\usepackage{amssymb}	

\usepackage{epsfig}
\usepackage{mathrsfs}
\usepackage{relsize}
\usepackage{mathtools}
\usepackage{multirow}
\usepackage{float}

\newcommand{\eps}{\mathrm{\epsilon_{ff}}}
\newcommand{\msun}{\mathrm{M}_\odot}
\newcommand{\lsun}{\mathrm{L}_\odot}
\newcommand{\lhcn}{L_\mathrm{HCN}}
\newcommand{\alphaHCN}{\alpha_{\rm HCN}}
\newcommand{\lir}{L_\mathrm{IR}}
\newcommand{\nh}{n_\mathrm{H}}
\newcommand{\mh}{m_\mathrm{H}}

\newcommand{\mach}{\mathcal{M}}
\newcommand{\dvdr}{dv/dr}
\newcommand{\ratio}{X_{\rm HCN}}
\newcommand{\perc}{\mathrm{\%}}
\newcommand{\amp}{\mathrm{\&}}

\newcommand{\km}{\,\mathrm{km}}
\newcommand{\K}{\,\rm K}
\newcommand{\pc}{\,\mathrm{pc}}

\newcommand{\Myr}{\,\mathrm{Myr}}
\newcommand{\Gauss}{\mathrm{G}}
\newcommand{\cm}{\,\mbox{cm}}
\newcommand{\g}{\,\mbox{g}}

\newcommand{\divergence}{\nabla \cdot \mathbf{v}}

\newcommand{\kms}{\km\,\mathrm{s}^{-1}}

\newcommand{\gcm}{\g\cm^{-3}}
\newcommand{\funnyunits}{\mathrm{K\,km\,s^{-1}\,pc^2}}
\newcommand{\sfrunits}{\msun\,\mathrm{yr^{-1}}}
\newcommand{\sfrhcnunits}{\sfrunits/\,(\funnyunits)}

\usepackage{color}
\usepackage[normalem]{ulem}

\definecolor{darkgreen}{rgb}{0.13, 0.55, 0.13}

\title[HCN--SFR Calibration]{Numerical Calibration of the HCN--Star Formation Correlation}
\author[Adam Onus]{
Adam~Onus$^{1}$\thanks{E-mail: u6380265@anu.edu.au}, Mark~R. Krumholz$^{1}$, Christoph~Federrath$^{1}$\\
$^{1}$Research School of Astronomy and Astrophysics, The Australian National University, Canberra, ACT~2611, Australia}

\date{\today}
\pubyear{2017}

\begin{document}
\label{firstpage}
\pagerange{\pageref{firstpage}--\pageref{lastpage}}
\maketitle

\begin{abstract}
HCN(1--0) emission traces dense gas and correlates very strongly with star formation rates (SFRs) on scales from small Milky Way clouds to whole galaxies. The observed correlation offers strong constraints on the efficiency of star formation in dense gas, but quantitative interpretation of this constraint requires a mapping from HCN emission to gas mass and density. In this paper we provide the required calibration by post-processing high-resolution simulations of dense, star-forming clouds to calculate their HCN emission ($\lhcn$) and to determine how that emission is related to the underlying gas density distribution and star formation efficiency. We find that HCN emission traces gas with a luminosity-weighted mean number density of $0.8-1.7 \times 10^4 \cm^{-3}$ and that HCN luminosity is related to mass of dense gas of $\gtrsim 10^4 \cm^{-3}$ with a conversion factor of $\alphaHCN \approx 14\,\msun/\,(\funnyunits)$. We also measure a new empirical relationship between the star formation rate per global mean freefall time ($\eps$) and the SFR--HCN relationship, SFR/$\lhcn \approx 2.0 \times 10^{-7}\,(\eps/0.01)^{1.1}\,\sfrhcnunits$.  The observed SFR--HCN correlation constrains $\eps\approx 1\perc$ with a factor of $\sim 3$ systematic uncertainty. The scatter in $\eps$ from cloud to cloud within the Milky Way is a factor of a few. We conclude that $\lhcn$ is an effective tracer of dense gas and that the IR--HCN correlation is a significant diagnostic of the microphysics of star formation in dense gas.
\end{abstract}



\section{Introduction}

The HCN(1--0) line is one of the brightest molecular lines produced in most star-forming galaxies, and it has a much higher critical density \citep[$2-20\times10^5\cm^{-3}$, see][]{Shirley15a,LineRatios,J-D17} than the brighter lines of CO \citep[$\sim$$10^2\cm^{-3}$, e.g.,][]{LineRatios}. It is thought to trace gas at number densities $\nh \gtrsim 6 \times 10^4 \cm^{-3}$ typically associated with active star formation. Consequently, HCN emission is of great interest and has been extensively studied over the past two decades both observationally \citep[e.g.,][]{G&S1,G&S2,wu05,wu,G-B,Kepley,Usero,Chen15a,Bigiel15,Bigiel} and theoretically \citep[e.g.,][]{SlowSFR,OtherKrumholz07,Narayanan08b,Hopkins2013,LineRatios}. HCN is a particularly useful tool because its high critical density means that HCN emission provides constraints on the volume density of the emitting gas, while lower critical density tracers such as CO are sensitive primarily to total mass, and offer little constraint on volumetric properties. Extragalactic observations of HCN provide one of the few methods available to study dense, star forming clumps in external galaxies, which are too small to resolve spatially, but which can be separated from their larger-scale environments because they are much brighter in HCN emission. Indeed, the opportunity offered by comparing Galactic and extragalatic HCN emission has motivated several studies of HCN emission in the Milky Way in order to provide a comparison sample for extragalactic surveys \citep[e.g.,][]{HCNHCO,wu05,wu,HCNtracer,Stephens16a}.

The key result of HCN studies to date is that HCN(1--0) luminosities correlate very strongly with star formation rates (SFRs) both in the Milky Way \citep{HCNHCO,wu05,wu,HCNtracer,Stephens16a} and in extragalactic observations \citep{G&S1,G&S2,G-B,Kepley,Usero,Chen15a,Bigiel15,Bigiel}. This correlation is close to but not exactly linear, and extends over many order of magnitude in HCN luminosity and SFR. To the extent that HCN emission provides a direct measurement of the mass of gas at a particular density, this correlation can be used to constrain the local efficiency of star formation, $\eps$, the fraction of gas converted into stars per freefall time \citep{Mark05,WhatIOriginallyWantedToDo}. Values of $\eps$ are theoretically significant because they directly relate to physical parameters of cloud structure and to the nature of star formation \citep{P&N,Mark05,H&C,WhatIOriginallyWantedToDo,C&MBondi,Hopkins2013}. Moreover, because $\eps$ is a scale-free quantity, it can be measured in objects of very different physical scales, enabling comparisons of star formation efficiency across scale.

There are many models for $\eps$ which lack calibration and constraint. Observations of $\eps$ based on direct measurements of individual clouds in the Milky Way or nearby galaxies have for the most part indicated uniformly low values of $\eps\approx 1\%$ \citep{Krumholz12a,Evans14a,Vutisalchavakul16a,Heyer16a,LeroyM51}, though there are a few exceptions \citep{Murray11b, Lee16a}. Some authors have proposed that $\eps$ has a small average value because it is negligible at densities too low to be traced by HCN emission but rises significantly in dense gas ($\nh \gtrsim 6 \times 10^4 \cm^{-3}$) traced by HCN \citep[e.g.,][]{SFRinGMC, Lada12a, Shimajiri17a}. Other models predict that star formation is fast and efficient only in all collapsed structures \citep{H&H,Z-A&V-S}, and occurs slowly or not at all in gas that is not self-gravitating. These models predict $\eps$ to be low for gas traced by CO emission (which has lower density) but is high in gas traced by HCN emission (high densities). In contrast, other models predict small values of $\eps$ independent of density \citep[e.g.][]{Mark05,Padoan11a,WhatIOriginallyWantedToDo}. In principle all of these models, and many others, can be constrained by the value of $\eps$ in dense gas as traced by the IR--HCN correlation \citep{SlowSFR,OtherKrumholz07,Hopkins2013,LineRatios}.

However, quantitative interpretation of the IR--HCN correlation is hampered by uncertainty about the exact density probed by the HCN(1--0) line, and by the fact that the conversion from masses above this density to HCN emission ($\alphaHCN$) is only approximately known. Published estimates for these quantities thus far have been based solely on models using idealised clouds or density distributions \citep[e.g.,][]{OtherKrumholz07,LineRatios}. The relationship between HCN emission, density, and star formation has yet to be calibrated by detailed simulations that resolve turbulent structure in the emitting gas, whilst self-consistently computing star formation. The published work that has come closest to attempting such a calculation is \citet{Hopkins2013}, but their simulations only barely resolve densities where HCN emission is strong, only measure gas mass above a density threshold rather than calculating HCN emission directly, and treat star formation via a sub-grid model rather than resolving gravitational collapse to individual stars directly, so $\eps$ is an input rather than an output of the simulation.

Here we address this omission in the literature using high-resolution simulations that self-consistently compute SFR and $\eps$. We post-process these simulations to self-consistently calculate the HCN luminosity and its relationship to the gas density distribution. We then use the result of these efforts to calibrate the value of $\alphaHCN$ and the HCN--density dependence and to determine how SFR, $\lhcn$ and $\eps$ are correlated.

\autoref{sec:method} summarises the numerical method of our simulations, including how we incorporate HCN luminosity models into the data. Our results are presented in \autoref{sec:results}, where we find that HCN emission is indeed distributed over regions of higher density in our simulations and also define an empirical relation between SFR/$\lhcn$ and $\eps$. In \autoref{sec:observation} we review existing literature and compare our simulations to observations with similar characteristics, using our results to interpret this observed data. We summarise our findings and conclusions in \autoref{sec:conclusion}.

\section{Computing the HCN Luminosity} \label{sec:method}

\begin{table*}
\caption{Key simulation parameters}
\label{tab:sims}
\def\arraystretch{1.1}
\setlength{\tabcolsep}{3.0pt}
\begin{tabular}{| l c c c c c c c c c c c |}
\hline
\hline
Simulation & Turbulence & $\sigma_\mathrm{v}\,(\mathrm{km s^{-1}})$ & $\mach$ & $B \, (\mu\mathrm{G})$ & $\beta$ & $\mach_\mathrm{A}$ & Jet+Radiation Feedback & $N_{\mathrm{res}}^3$ & SFR$\,(\sfrunits)$ & $\eps$ & $\lhcn \, (\funnyunits$) \\
(1) & (2) & (3) & (4) & (5) & (6) & (7) & (8) & (9) & (10) & (11) & (12) \\
\hline
\hline
G            & None & $0$ & $0$ & $0$ & $\infty$ & $\infty$ & No  & $1024^3$ & $1.6\!\times\!10^{-4}$ & $0.47$ & $4.6$ \\
GT      & Mix & $1.0$ & $5.0$ & $0$ & $\infty$ & $\infty$ & No  & $1024^3$ & $8.3\!\times\!10^{-5}$ & $0.25$ & $17$ \\
GTB   & Mix & $1.0$ & $5.0$ & $10$ & $0.33$ & $2.0$ & No  & $1024^3$ & $2.8\!\times\!10^{-5}$ & $0.083$ & $14$ \\
GTBJR & Mix & $1.0$ & $5.0$ & $10$ & $0.33$ & $2.0$ & Yes & $2048^3$ & $1.0\!\times\!10^{-5}$ & $0.031$ & $13$ \\
\hline
\hline
\end{tabular}
\begin{minipage}{\linewidth}
\textbf{Notes.} Column 1: simulation name. Columns 2--4: the type of turbulence driving, turbulent velocity dispersion, and turbulent rms sonic Mach number. Columns 5--7: magnetic field strength, the ratio of thermal to magnetic pressure (plasma $\beta$), and the Alfv\'en Mach number. Column 8: whether jet/outflow feedback and radiation was included or not. Column 9: maximum grid resolution. Columns 10--11: absolute SFR and the SFR per mean global freefall time. Column 12: the total HCN luminosity at SFE of $5\perc$. Simulations are listed in order of increasing physical complexity.
\end{minipage}
\end{table*}

\subsection{Numerical Simulations} \label{sec:simulationmethod}

\subsubsection{Simulation methods}

We use high-resolution simulations from \citet{Turbulence}, and we refer readers to that paper for full details on the computational setup. Here we only summarise the most important features. The simulations solve the equations of compressible magnetohydrodynamics through use of the multi-physics, adaptive mesh refinement \citep{BCmethod} code FLASH (v4) \citep{flash1,flash2} in conjunction with the positive--definite HLL5R Riemann solver \citep{waagan}. These simulations include turbulence generated by an Ornstein--Uhlenbeck process \citep{Esw-Pope,Schmidty} that naturally generates a mixture of solenoidal and compressible modes with a driving parameter $b = 0.4$ \citep{Christoph2010b}.

All simulations are periodic boxes of size $L = 2\pc$, total cloud mass $M = 388\,\msun$ and a mean density $\rho_0=3.28\times\mathrm{10^{-21} \gcm}$, corresponding to a global mean freefall time of $t_\mathrm{ff}=1.16\Myr$. We have four simulations of increasing physical complexity. Simulation G includes only gas self-gravity, with no turbulence driving or magnetic fields. GT includes self-gravity and driven hydrodynamic turbulence. GTB adds magnetic fields, and GTBJR includes protostellar jet and radiation feedback as well \citep[following the implementation described by][]{Jets,feedback}. Each simulation has an initial virial ratio $\alpha_\mathrm{vir} = 1.0$; those with magnetic fields have a plasma beta of $\beta = 0.33$ (corresponding to an Alfven Mach number $\mach_\mathrm{A} = 2.0$). Simulations including turbulence have velocity dispersion of $\sigma_\mathrm{v} = 1\kms$ and an rms Mach number of $\mach = 5$, resulting from a sound speed of $c_\mathrm{s} = 0.2\kms$ at temperature $T = 10\K$. Simulations with a magnetic field initially have a uniform field of $B = 10\,\mu\Gauss$ which is subsequently compressed, tangled and twisted by the turbulence. These properties are summarised in Columns 3 -- 7 of \autoref{tab:sims}.

We measure the star formation rate (SFR) in the simulations through the sink particle method developed by \citet{Christoph2010a}, which is enhanced by applying a jet feedback module \citep{Jets}. The simulation's SFRs span an order of magnitude, which gives us an advantageous calibration set which can be compared to observations to see which simulations match the observed SFR--$\lhcn$ relation.

\subsubsection{Uncertainties in the simulations}
The main uncertainties in the numerical simulations are related to the choice of boundary conditions and the absence of chemical evolution and associated detailed heating and cooling effects through radiative transfer. Here we briefly discuss potential limitations resulting from these approximations.

The simulations use periodic boundary conditions. This choice approximates the effects of the surrounding large-scale gas (flows and gravity) on the cloud scales modelled -- here a $(2\,\mathrm{pc})^3$ section of a molecular cloud. Although real clouds are not periodic, the alternative choice (assuming that the cloud is isolated) is also not realistic. A full galaxy simulation producing clouds self-consistently and then zooming into those clouds would be necessary to improve on the boundary conditions.

The simulations follow a polytropic equation of state \citep[see Eqs.~3 and 4 in][]{Jets} to approximate the thermodynamical evolution during star formation from low-density molecular gas to stellar densities. The simulations themselves do not include detailed non-equilibrium chemical evolution, or heating/cooling through direct radiative transfer. However, the GTBJR simulation does include a simple radiative feedback approximation \citep{feedback}. These limitations may affect some of the details related to where and in what excitation state HCN should be expected to form and to be observable in the simulations. However, we explore the effects of varying the temperature in the post-processing with DESPOTIC (see next section and last two models in Table~\ref{tab:Equation_Parameters}) and find that this introduces uncertainties in our main results by only $\sim30\%$, while the assumed HCN abundance leads to uncertainties by a factor of $\sim2$.

The main purpose of the simulations is to provide a set of models with intrinsically varying $\eps$ that we can use to produce realistic-looking HCN mock observations. This is achieved with the present set of simulations, as they cover a factor of 15 in $\eps$ and include most of the relevant physical effects that control the star formation rate and structure of molecular clouds \citep[gravity, turbulence, magnetic fields, jet and radiative feedback; for details, see][]{Turbulence}.

\subsection{Modeling HCN Emission} \label{sec:HCNmethod}

We use the code DESPOTIC \citep{DESPOTIC} to calculate the HCN luminosity of every cell in the simulations. DESPOTIC solves the equations of statistical equilibrium for the HCN level population, including non-local thermodynamic equilibrium (LTE) effects. It treats optical depth effects using an escape probability formalism, and for the purposes of this paper we estimate the escape probabilities using the large velocity gradient (LVG) approximation \citep{Goldreich74a,LVGapprox}. We refer readers to \citet{DESPOTIC} for full details of the model and numerical method. For all calculations we use molecular data from the Leiden Atomic and Molecular Database \citep{lambda}\footnote{\url{http://home.strw.leidenuniv.nl/~moldata/}}; the underlying collision rate data for HCN are from \citet{Dumouchel10a} and for CO (see below) are from \citet{Yang10a}. We assume that the gas is molecular hydrogen plus helium in the usual cosmic ratio of 25\% He by mass, and that the $\mathrm{H_2}$ has an ortho-to-para ratio (OPR) of 0.25, consistent with typical values of cold cores and shocks \citep[e.g., see the recent review by][]{Wakelam17}. The choice of OPR will not affect the results substantially, since the \citet{Dumouchel10a} collision rate coefficients we use for the excitation of HCN by $\mathrm{H_2}$ do not distinguish between ortho- and para- forms, and in this case DESPOTIC assumes the rates are identical. The OPR only becomes relevant when DESPOTIC computes temperature self-consistently (see below), because the temperature is primarily controlled by CO emission, and the \citet{Yang10a} rates for collisional excitation of CO do distinguish between para-$\mathrm{H_2}$ and ortho-$\mathrm{H_2}$.

\begin{table}
\caption{Key Parameters of HCN(1--0) Emission Models
\label{tab:models}
}
\def\arraystretch{1.0}
\setlength{\tabcolsep}{10.7pt}
\begin{tabular}{| l c c c |}
\hline
\hline
Model Name & $\ratio$ & $T$ \,(K) & $dv/dr$ \\
(1) & (2) & (3) & (4) \\
\hline
\hline
Standard & $1.0 \times 10^{-8}$ & 10 & $\divergence$\\
LOS & $1.0 \times 10^{-8}$ & 10 & Line-of-Sight\\
Low HCN & $3.3 \times 10^{-9}$ & 10 & $\divergence$\\
High HCN & $3.0 \times 10^{-8}$ & 10 & $\divergence$\\
High Temp & $1.0 \times 10^{-8}$ & 20 & $\divergence$\\
Varied Temp & $1.0 \times 10^{-8}$ & Varied & $\divergence$\\
\hline
\hline
\end{tabular}
\begin{minipage}{\linewidth}
\textbf{Notes.} Column 1: model name. Column 2: HCN abundance $\ratio \equiv n_{\rm HCN}/n_{\rm H}$. Column 3: gas temperature; see main text for details of the Varied Temp run. Column 4: method used to approximate $dv/dr$ in the LVG optical depth (see main text): velocity divergence $\divergence$ or an x--axis line-of-sight velocity.
\end{minipage}
\end{table}

We present six different models of HCN emission, chosen to bracket our uncertainties on quantities such as the HCN abundance and gas temperature. We summarise the features of these models in \autoref{tab:models}. For our fiducial model, denoted ``Standard" in \autoref{tab:models}, we assume an abundance ratio of $\ratio \equiv n_\mathrm{HCN}/n_\mathrm{H} = 10^{-8}$ \citep{Tieftrunk98a} with a constant gas temperature of $10\K$, and we take the velocity gradient $dv/dr$ that enters the LVG optical depth to be $\divergence$, where $\mathbf{v}$ is the velocity field in the simulations\footnote{One is required to choose an approximation for $dv/dr$ because the LVG approximation is one-dimensional, and thus there is some ambiguity in how to apply it to our three-dimensional simulations. The line luminosity escaping to an observer is most directly connected to the gradient in the line of sight velocity, while the radiative trapping factor that enters into the level populations is sensitive to the average of the velocity gradient over all directions, which is more closely related to $\divergence$.}. Our second model is identical to the first, except that we estimate the optical depth using the line-of-sight velocity gradient. Our third and fourth models differ from the fiducial one in that they use HCN abundances that are a factor of three lower and higher, respectively. This roughly spans the plausible range of HCN abundance in the dense ISM for gas of near-Solar metallicity \citep[e.g.][]{Gracia-Carpio08a,Meier14a,Vollmer17a}. The fifth model assumes a higher gas temperature of $20\K$, but is otherwise identical to the fiducial case. The sixth and final model, rather than using a fixed gas temperature, instead uses a gas temperature computed using DESPOTIC's thermal equilibrium calculation routine, whilst assuming $\ratio = 10^{-8}$ and using $\divergence$ for the velocity gradient as in the first model. For the purposes of the temperature calculation we include cosmic ray and photoelectric heating, cooling by $^{12}$CO and $^{13}$CO line emission, and dust--gas thermal energy exchange. We adopt a primary ionisation rate of $10^{-16}$ H$^{-1}$ s$^{-1}$ \citep[e.g.,][]{Indriolo12a}, a far ultraviolet radiation intensity ten times the Solar neighbourhood value ($\chi=10$ in DESPOTIC's notation), a $^{12}$CO abundance of $n_{\rm CO}/\nh = 10^{-4}$, and $^{13}$CO abundance of $n_{\rm ^{13}CO}/\nh = 5.0\times 10^{-7}$. All other parameters use DESPOTIC's default values -- see \citet{DESPOTIC} for details. The resulting gas temperatures are for the most part in the range $10-20\K$, though they can reach as high as $\sim$$30\K$ and as low as $\sim$$5\K$ for the cells with the smallest and largest velocity gradients at densities too low for significant dust coupling.

For all six cases we use DESPOTIC to generate a table of HCN luminosities per H nucleus, $\lhcn/n_{\rm H}$, as a function of gas number density $n_{\rm H}$ from $10^2 - 10^{10} \cm^{-3}$ and velocity gradient $\dvdr$ from $10^{-2} - 10^2 \kms\pc^{-1}$. We generate HCN luminosities for each cell by interpolating in $\log (\nh)$ and $\log (\dvdr)$ with a two-dimensional cubic spline. We convert from the mass density $\rho$ in the simulation to number density assuming a standard cosmic abundance ratio of 1 Helium per 10 Hydrogen nuclei, giving a mean gas mass per free H nucleus $\mh = 2.34 \times 10^{-24}\g$. We apply the tabulated HCN luminosities to snapshots of our simulations using the software package yt \citep{yt}. Our simulation post-processing code is freely available at \url{http://bitbucket.org/aonus/hcn}.

\section{Results} \label{sec:results}

\begin{figure}
\centering
\includegraphics[width=1.1\linewidth]{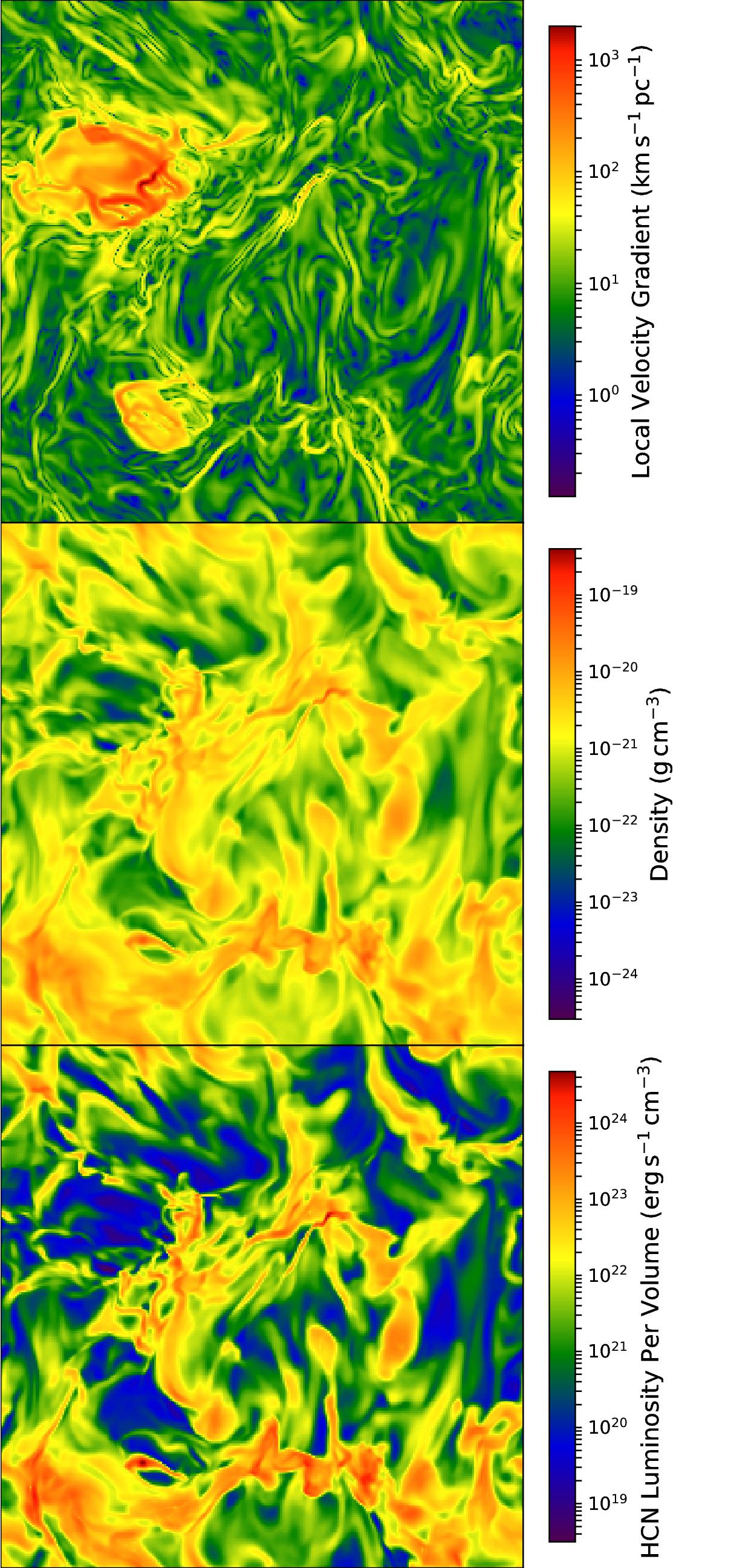}
	\vspace{-0.3cm}
\caption{Slice plots (each $2\pc\,\times\,2\pc$ in size) for simulation GTBJR at the time when the star formation efficiency is $5\perc$. In the top panel we plot the local velocity gradient $\divergence$, in the middle panel we plot density, and at the bottom we plot the corresponding HCN luminosity per unit volume for our Standard emission model.}
\label{fig:sliceplots}
\end{figure}

\begin{figure}
\centering\offinterlineskip
		\vspace*{-0.6cm}
		\includegraphics[width=\linewidth]{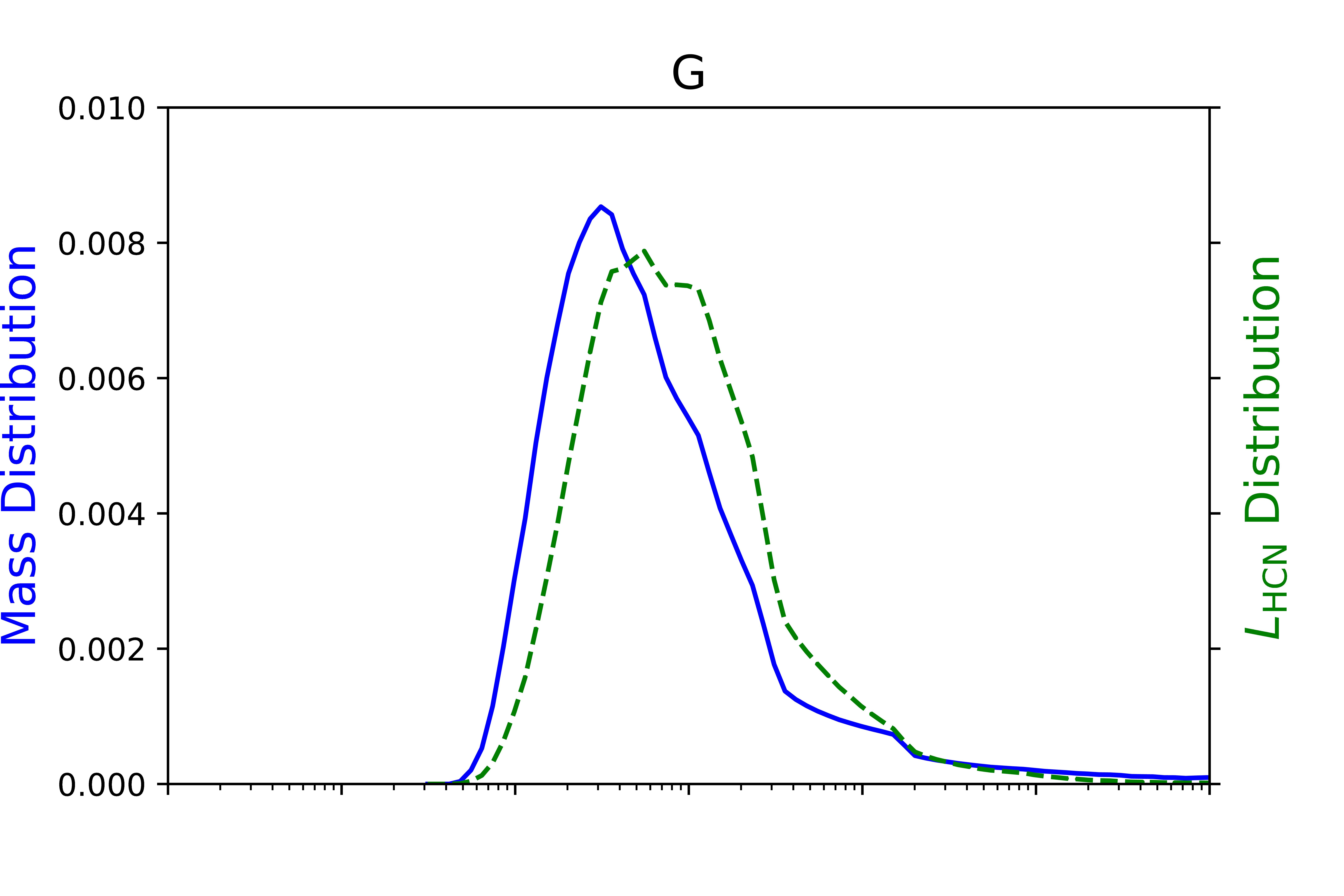} \\
        \vspace*{-0.6cm}
        \includegraphics[width=\linewidth]{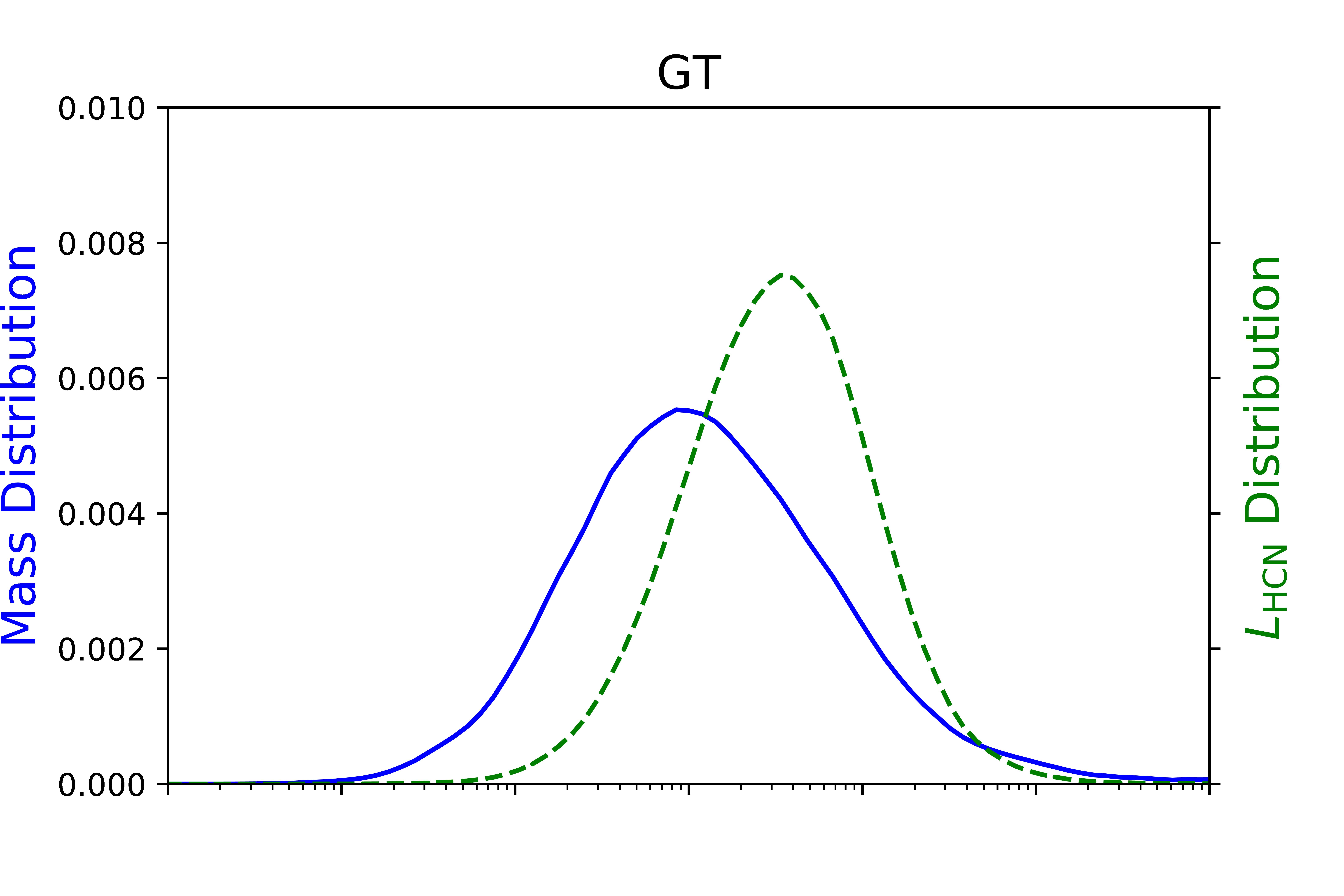} \\
        \vspace*{-0.6cm}
		\includegraphics[width=\linewidth]{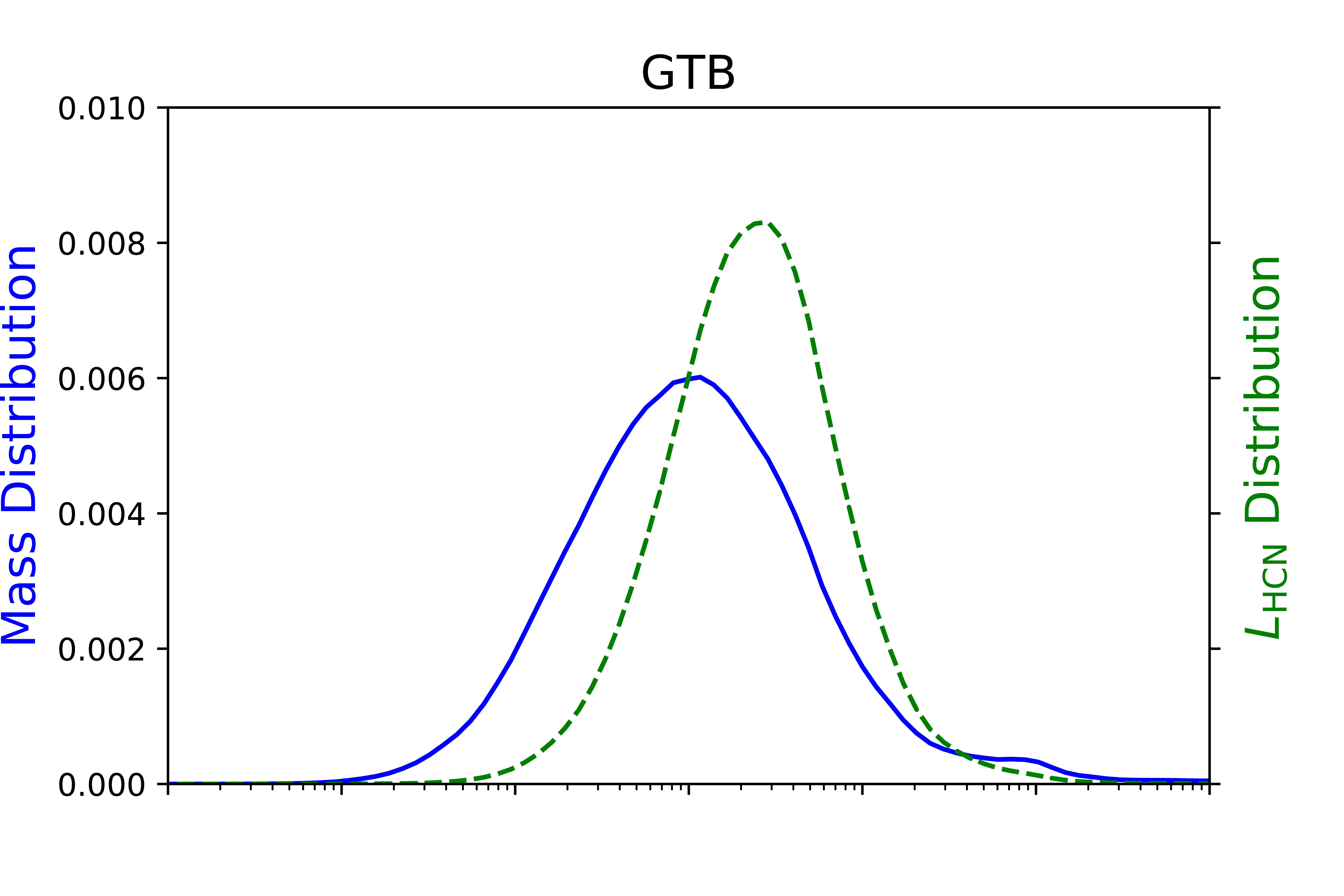} \\
        \vspace*{-0.6cm}
		\hspace{-0.4cm} \includegraphics[width=1.03\linewidth]{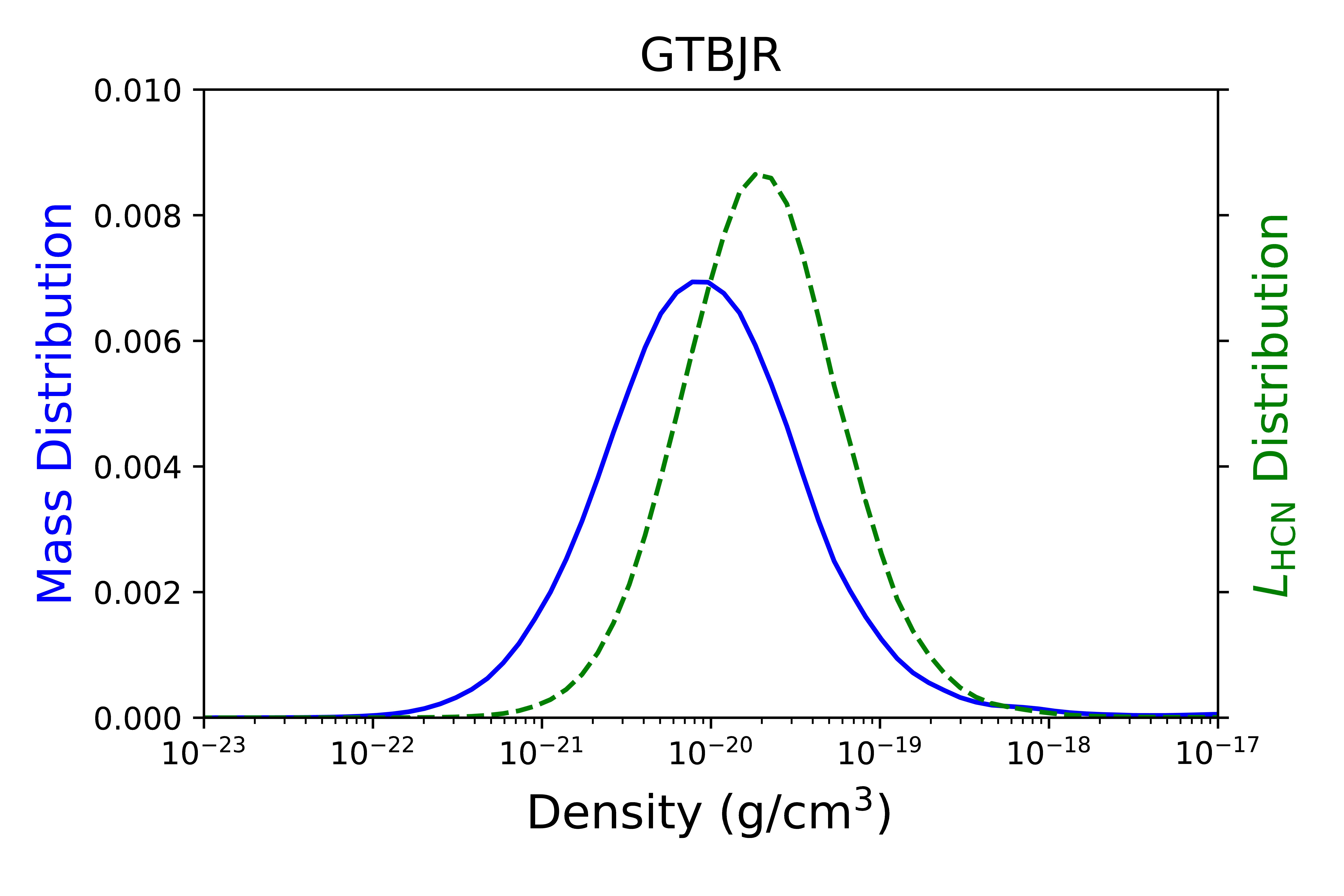} \\
        \vspace{-0.3cm}
\caption{PDFs of the density distributions with respect to cloud mass (in solid blue) and HCN luminosity (Standard model, in dashed green) for each of our simulations (at SFE of 5$\perc$): Gravity only (G, top panel), Gravity + Turbulence (GT, second panel), Gravity + Turbulence $\amp$ Magnetic Fields (GTB, third panel) and Gravity + Turbulence $\amp$ Magnetic $\amp$ Jet Feedback $\amp$ Radiation (GTBJR, bottom panel).}
\label{fig:densities}
\end{figure}

\subsection{What Density Range Does HCN Emission Trace?} \label{sec:profiles}

\autoref{fig:sliceplots} shows the distribution of HCN luminosity (bottom panel) in comparison to density (middle panel) and velocity gradient (top panel) for the Standard model in a slice through the GTBJR simulation at a star formation efficiency, SFE $\equiv \mathrm{M_{stars}/(M_{stars} + M_{gas})}$, of $5\perc$. We can observe a clear correlation between the density distribution and HCN luminosity. That is, regions of denser gas (shown in red) correspond to regions of high $\lhcn$ and likewise regions of low density (shown in blue) correspond to regions of low $\lhcn$. However, this correlation is predominantly in high-density regions. In low-density regions, the HCN luminosity drops much faster than the density, resulting in a considerably larger dynamic range of $\lhcn$ than density. This supports the idea of HCN as a dense gas tracer. In contrast, we can see no clear correlation between $\dvdr$ and $\lhcn$, which is indicative of the subtlety of the emissivity effect of the velocity gradient compared to density. Regions of high $\dvdr$ tend to correspond to regions of low density (and so $\lhcn$ is dominated by the effects of density), otherwise the $\dvdr$ lends minimal character to $\lhcn$ due to a small dynamic range.

In \autoref{fig:densities}, we plot the probability distribution functions (PDFs) for total mass and HCN luminosity with respect to density in each simulation at the time when the SFE is $5\perc$. The mass PDFs are well-approximated by log-normal distributions, as expected \citep{P&N,H&C,WhatIOriginallyWantedToDo}. The exception is the Gravity only simulation (G, top panel), in which we observe an extended power-law tail at high density. This abnormality can be attributed to a large $\eps$ of 0.47, as the power-law tails arise as a result of strong gravitational collapse \citep{Klessen00,FedKless13}

In the three other simulations, we observe the general trend that the HCN luminosity distribution is always centered around a greater average density and is less broadly distributed than the cloud mass PDF. The luminosity PDF peaks in the range $2\times10^{-20} - 4\times10^{-20}\gcm$, which corresponds to a number density of $0.8\times10^4 - 1.7\times10^4\cm^{-3}$. This is a factor of $\sim$$5$ less than what is assumed in studies such as \citet{G&S2}, and at the low end of the range suggested in other observational studies \citep{Usero}. However, mass is distributed with a mean density of $\sim 8 \times 10^{-21}\gcm$ ($\sim 3.4 \times 10^3\cm^{-3}$), so we still find that HCN emission traces gas at densities $2.5 - 5$ times greater than the mean density in the simulations.

In \autoref{tab:alphaHCN} we present the conversion factor between $\lhcn$ and mass, $\alphaHCN$, for each simulation with each emission model. We compare the conversion for gas above the mean density for the luminosity distribution in our simulations ($\nh \approx 1.0\times10^4\cm^{-3}$) and above the predicted high density threshold for HCN(1--0) emission, $\nh \approx 6.0\times10^4\cm^{-3}$ \citep{G&S1,LineRatios}. We find $\alphaHCN = 14 \pm 6\,\msun/\,(\funnyunits)$, where we quote the mean for the Standard emission model plus or minus the standard deviation of each model for each simulation (excluding G, with $\nh = 1.0\times10^4\cm^{-3}$). $\alphaHCN$ is thought to range between $3-30\,\msun/\,(\funnyunits)$ based on various estimates of observed values \citep{G&S1,wu05,SlowSFR,Shimajiri17a}. This is typically supported by our results irrespective of the threshold density, albeit weighted towards larger values (with exception to the G simulation, which is not very realistic anyway). $\alphaHCN$ calculated with our mean density threshold is very similar to observed averages \citep{wu05,SlowSFR} of $\sim 10\,\msun/\,(\funnyunits)$, and is a factor of $1.5-2$ less than when calculated with the high density threshold of HCN(1--0) emission. This suggests that previous overestimates of densities traced by HCN emission ($\nh \gtrsim 6 \times10^4\cm^{-3}$) do not accurately reflect the true conversion between mass and luminosity for dense gas, as well as giving underestimates of $t_\mathrm{ff}$ and similar values. Our findings are also consistent with other suggestions in the literature that a significant portion of the total HCN emission comes from gas with densities up to a factor of $\sim 10$ below the critical density \citep[e.g.][]{Shirley15a, Shimajiri17a}.

\begin{table*}
\caption{$\alphaHCN$ for each emission model and simulation}
\label{tab:alphaHCN}
\def\arraystretch{1.0}
\setlength{\tabcolsep}{9.6pt}
\begin{tabular}{| l c c c c c c c |}
\hline
\hline
Simulation & Threshold Density\,($\rm cm^{-3}$) & Standard & LOS & Low HCN & High HCN & High Temp & Varied Temp \\
(1) & (2) & (3) & (4) & (5) & (6) & (7) & (8) \\
\hline
\hline
G           & $1.0 \times 10^4$ & 63 & 48 & 83 & 53 & 37 & 41 \\
            & $6.0 \times 10^4$ & 120 & 73 & 130 & 110 & 55 & 58 \\
            \hline
GT          & $1.0 \times 10^4$ & 12 & 13 & 23 & 7.8 & 8.9 & 12 \\
            & $6.0 \times 10^4$ & 19 & 21 & 25 & 16 & 11 & 13 \\
            \hline
GTB         & $1.0 \times 10^4$ & 16 & 17 & 29 & 10 & 11 & 15 \\
            & $6.0 \times 10^4$ & 31 & 32 & 38 & 28 & 17 & 19 \\
            \hline
GTBJR       & $1.0 \times 10^4$ & 15 & 15 & 28 & 9.5 & 11 & 14 \\
            & $6.0 \times 10^4$ & 25 & 25 & 32 & 23 & 14 & 16 \\
\hline
\hline
\end{tabular}
\begin{minipage}{\linewidth}
\textbf{Notes.} Column 1: simulation name. Column 2: Minimum density for which $\alphaHCN$ is measured. Columns 3--8: $\alphaHCN$ of each model in $\msun/\,(\funnyunits)$
\end{minipage}
\end{table*}

\subsection{Star Formation -- HCN Luminosity Ratio} \label{sec:SFRHCN}

\begin{figure}
\centering
\includegraphics[width=1.05\linewidth]{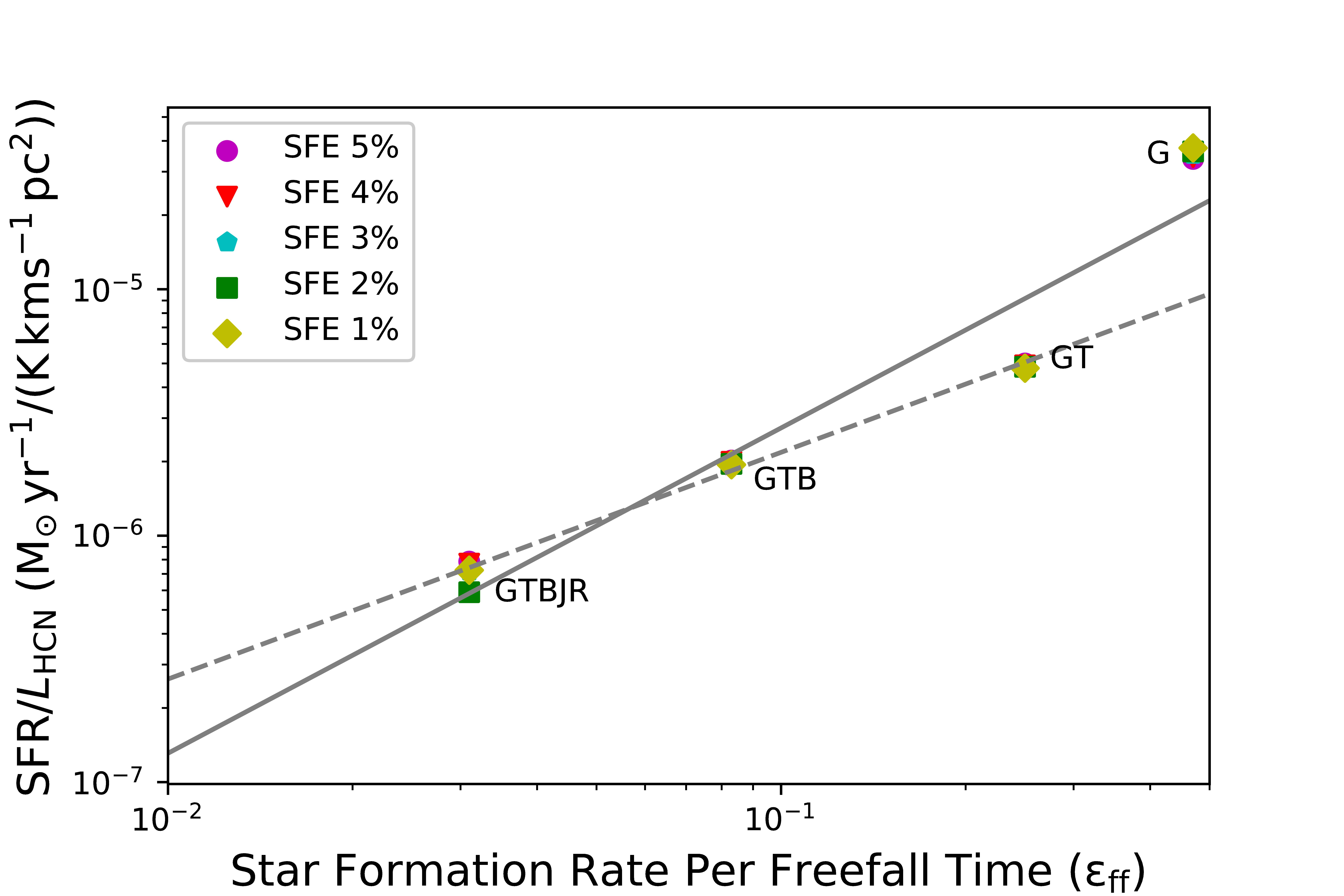}
	\vspace{-0.3cm}
\caption{Ratio of SFR/$\lhcn$ versus $\eps$ for all simulations at star formation efficiencies of $1-5\perc$, as indicated in the legend, using our Standard HCN emission model. The lines are linear least-squares fits to the simulation results when including the G simulation in calculations (solid) and when excluding the simulation (dashed) (where $\lhcn$ is averaged over all SFEs), using the parameters shown in \autoref{eq:equation}and \autoref{eq:equation2}.}
\label{fig:epsilon_ratio}
\end{figure}

\autoref{fig:epsilon_ratio} shows the ratio of SFR/$\lhcn$ versus $\eps$ for the Standard HCN emission model in each of our simulations. To characterise the level of fluctuations in SFR/$\lhcn$ over time we show this relationship measured at SFEs of $1\perc$ , $2\perc$, $3\perc$, $4\perc$ and $5\perc$; in these calculations we use the time-averaged star formation rate (since all observational tracers of star formation are also time-averaged), but we use the instantaneous HCN luminosity for each simulation snapshot (although our regression line is derived from the average of these values). We see that SFR/$\lhcn$ varies by less than a factor of two over this range in SFE, and thus is quite stable. Moreover, there is a very clear relationship between the value of SFR/$\lhcn$ and $\eps$, which is well-fit by
\begin{equation}
\centering
\frac{\rm SFR}{\lhcn} = 1.3\times10^{-7}\,\left(\frac{\eps}{0.01}\right)^{1.3}  \sfrhcnunits
\label{eq:equation}
\end{equation}
when we include G in the calculations, and
\begin{equation}
\centering
\frac{\rm SFR}{\lhcn} = 2.6\times10^{-7}\,\left(\frac{\eps}{0.01}\right)^{0.9}  \sfrhcnunits
\label{eq:equation2}
\end{equation}
when we exclude G; as with our estimates of $\alpha_{\rm HCN}$, it is potentially more informative to exclude G, since its density PDF is quite different than those of all the other simulations. We show each fit line in \autoref{fig:epsilon_ratio}.

We can repeat this procedure for all our other HCN emission models, fitting functions of the form
\begin{equation}
\frac{\rm SFR}{\lhcn} = \left(\frac{\rm SFR}{\lhcn}\right)_{0.01} \left(\frac{\eps}{0.01}\right)^{p}.
\label{eq:parameter}
\end{equation}
In all cases we find fits comparable in quality to that shown in \autoref{fig:epsilon_ratio}, with best fit parameters as shown in \autoref{tab:Equation_Parameters}.

\begin{table}
\caption{Fit Parameters for SFR/$\lhcn$ versus $\eps$}
\label{tab:Equation_Parameters}
\def\arraystretch{1.0}
\setlength{\tabcolsep}{2.4pt}
\begin{tabular}{| l l c c c |}
\hline
\hline
Model & Calibration & $({\rm SFR}/\lhcn)_{0.01}$ & $p$ & $\epsilon_{\rm ff,Bigiel}$ \\
& including G & ($\sfrhcnunits$) & & ($\%$) \\
(1) & (2) & (3) & (4) & (5) \\
\hline
\hline
Standard & Yes & $1.3 \times 10^{-7}$ & 1.3 & 1.1 \\
 & No & $2.6 \times 10^{-7}$ & 0.92 & 0.51 \\
 \hline
LOS & Yes & $1.4 \times 10^{-7}$ & 1.3 & 1.0 \\
 & No & $2.6 \times 10^{-7}$ & 0.93 & 0.51 \\
 \hline
Low HCN & Yes & $3.1 \times 10^{-7}$ & 1.3 & 0.54 \\
 & No & $5.9 \times 10^{-7}$ & 0.90 & 0.20 \\
 \hline
High HCN & Yes & $6.0 \times 10^{-8}$ & 1.4 & 1.9 \\
 & No & $1.3 \times 10^{-8}$ & 0.94 & 1.1 \\
 \hline
High Temp & Yes & $1.0 \times 10^{-7}$ & 1.3 & 1.3 \\
 & No & $2.0 \times 10^{-7}$ & 0.91 & 0.67 \\
 \hline
Varied Temp & Yes & $1.7 \times 10^{-7}$ & 1.2 & 0.86 \\
 & No & $3.0 \times 10^{-7}$ & 0.90 & 0.43 \\
\hline
\hline
\end{tabular}
\begin{minipage}{\linewidth}
\textbf{Notes.} Column 1: model name. Column 2: whether equation is calibrated including or excluding G simulation. Column 3: constant for \autoref{eq:parameter}. Column 4: exponent in \autoref{eq:parameter}. Column 5: $\eps$ predicted for the SFR$-\lhcn$ correlation in \citet{Bigiel} (see \autoref{sec:observation} and \autoref{fig:wucomparison}).
\end{minipage}
\end{table}

Our results indicate that the changes in how we apply the LVG method (as explored in the LOS model) produce only $\sim 10\%$ shifts in the predicted relationship between $\mbox{SFR}/\lhcn$ and $\eps$. Changes in the gas temperature within the plausible range of $\sim 10-20$ K produce shifts at the $\sim 30\%$ level at most. The parameter to which the results of each fit are most sensitive is the HCN abundance, where factor of 3 changes in the assumed value induce factor of $\sim$$2$ changes in the normalisation of the IR--HCN correlation. While the dependence is sublinear (as expected, since the changes are partially canceled by optical depth effects), the uncertainty in HCN abundance still clearly dominates the overall uncertainty. This uncertainty is comparable in size to the uncertainty produced by the decision whether or not to include simulation G in our fits.

\section{Implications for the Interpretation of Observations} \label{sec:observation}

Our simulations span a considerable range in $\eps$ (and thus $\mbox{SFR}/\lhcn$), though as we shall see their range is systematically offset from the range covered by observed systems. \citet{Bigiel} and \citet{Usero} find that $\lir/\lhcn \approx 900\,\lsun/(\funnyunits)$ well approximates the IR--HCN correlation observed on all scales. This suggests that the observed $\mbox{SFR}/\lhcn$ provides a strong constraint on $\epsilon_{\rm ff}$ and thus on the physics that governs star formation. Since most observational studies of the SFR$-\lhcn$ correlation use infrared luminosity as their SFR tracer, in order to exploit this constraint we must translate our simulated SFRs to infrared luminosities. For this purpose we adopt a conversion \citep{LIRrelation}
\begin{equation}
\frac{\rm SFR}{\lir} = 1.5\times10^{-10}\,\sfrunits/\,\lsun.
\label{eq:IRratio}
\end{equation}
Using this conversion together with \autoref{eq:equation}, we can immediately translate the observed relation $\lir/\lhcn \approx 900\,\lsun/(\funnyunits)$ into a measurement of $\eps$. For our standard emission model, the observed IR--HCN ratio corresponds to $\eps = 1.1\perc$ when using our fit that includes simulation G, and $\eps = 0.51\perc$ using the fit that excludes it. For the other emission models (\autoref{tab:Equation_Parameters}) inferred $\eps$ values fall in the range $0.5\perc - 1.9\perc$ with G. This range of values becomes $0.2\perc - 1.1\perc$ when excluding G, which is a factor of $\sim 2$ lower. Thus our results imply $\eps \approx 1\perc$ with roughly a factor of $\sim 3$ uncertainty.

\begin{figure}
\centering
	\includegraphics[width=1.0\linewidth]{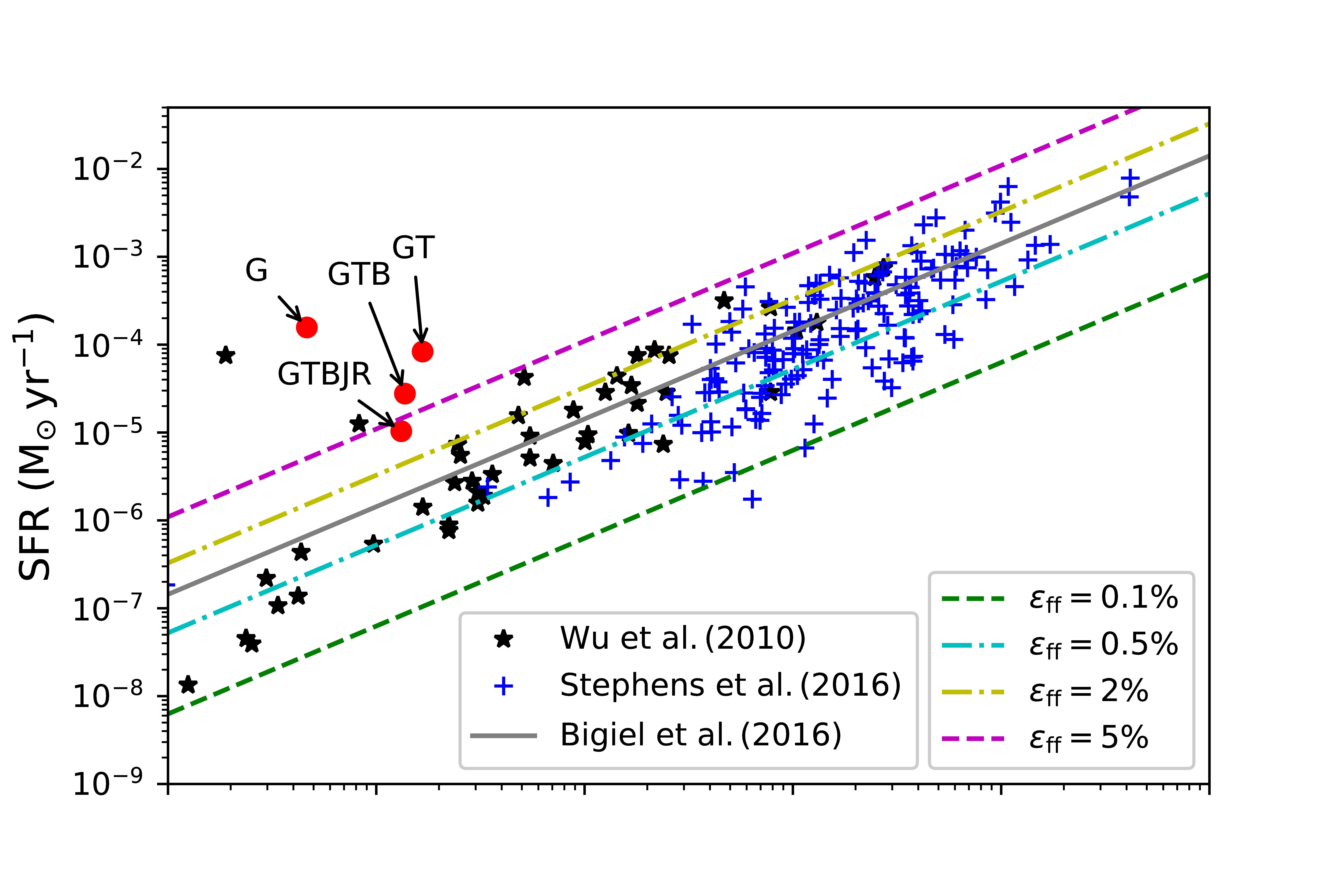}\\
        \vspace*{-0.65cm}
	\includegraphics[width=1\linewidth]{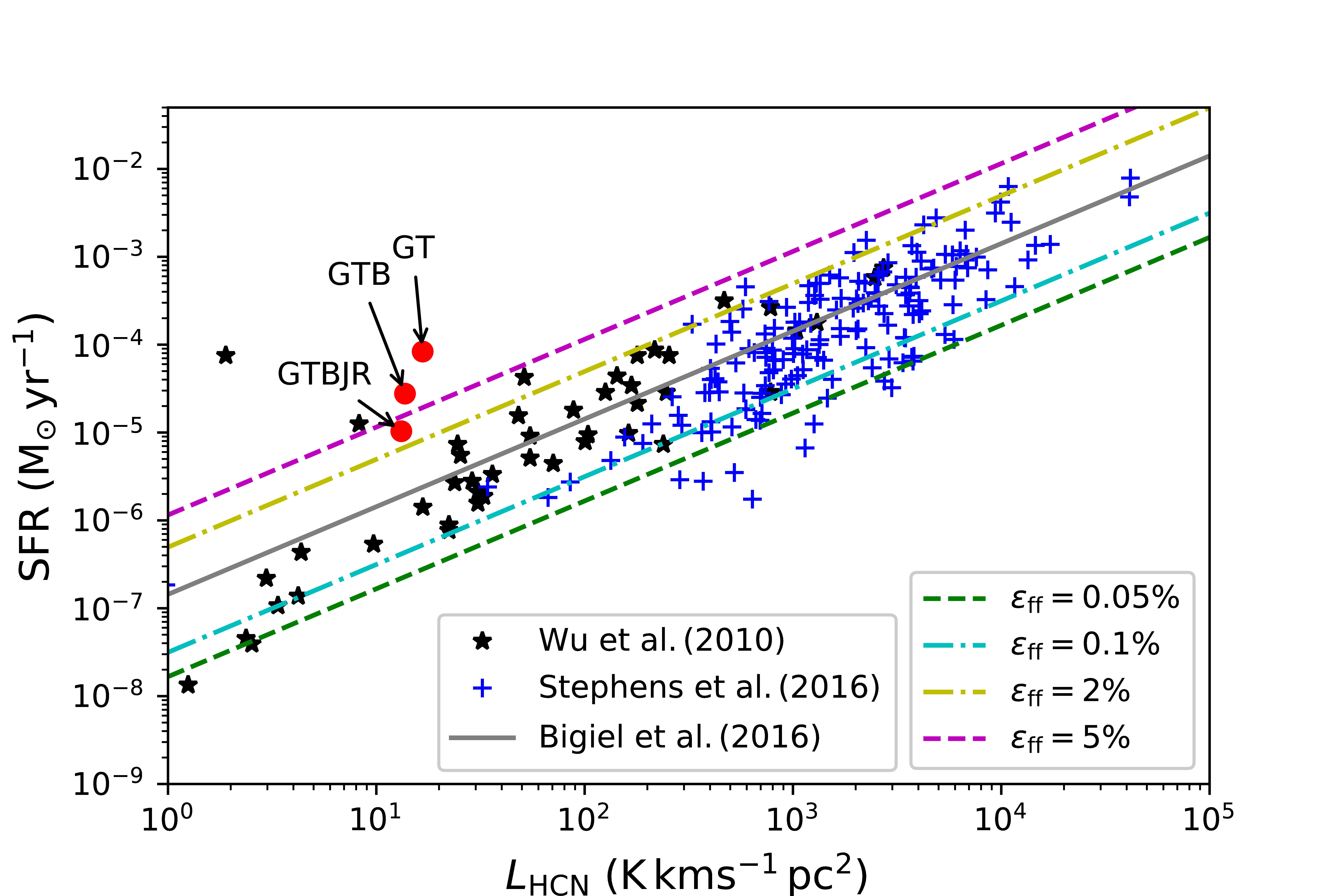}\\
\caption{SFR as a function of $\lhcn$. In the top panel we plot values when calibrating with G and in the bottom panel we plot values calibrating without G. We show observations of Milky Way sources from \citet{wu} (black stars) and \citet{Stephens16a} (blue +'s), as well as our simulations (red circles). The observations have been converted from $\lir$ to SFR using \autoref{eq:IRratio}. The solid gray line is the observed mean IR--HCN correlation from \citet{Bigiel}, which corresponds to $\eps = 1.1\perc$ when calibrating with G, or $\eps = 0.51\perc$ otherwise. The remaining lines show SFR/$\lhcn$ ratios for $\eps$ (green dashes: $0.1\perc$ (or $0.05\perc$ in the lower panel), cyan dot-dashes: $0.5\perc$ (or $0.1\perc$ in the lower panel), yellow dot-dashes: $2\perc$, magenta dashes: $5\perc$) as predicted by \autoref{eq:equation} for our standard emission model.}
\label{fig:wucomparison}
\end{figure}

In addition to interpreting the average IR--HCN relation in terms of $\eps$, our calibration allows us to do so on a source-by-source basis. In \autoref{fig:wucomparison} we overplot curves of constant $\eps$ for our standard model with observations of massive, dense gas clumps in the Milky Way from \citet{wu} and \citet{Stephens16a}; we also show our raw simulation results and the average relationship for comparison.

There are two immediate and obvious points to take from \autoref{fig:wucomparison}. The first is that, of our simulations, only the one with the lowest value of $\eps$ (simulation GTBJR) falls near the locus of observed points. Clearly simulations where star formation proceeds at high efficiency are strongly inconsistent with the observed IR--HCN relation. Indeed, even our simulation that forms stars least efficiently yields a value of $\eps$, or equivalently $\mbox{SFR}/L_{\rm HCN}$, that is near the upper envelope of the observed distribution. This is a symptom of the longstanding problem that simulations of star cluster formation (not simply the ones we use here) tend to produce stars too efficiently compared to observations. The origin of this discrepancy may lie in the lack of feedback from massive stars (  since the regions that \citet{Turbulence} simulates do not produce stars massive enough to drive H~\textsc{ii} regions or substantial winds), or in the fact that the simulations use a periodic box, and thus lack external forcing from ongoing accretion flows or cloud assembly. We refer readers to \citet{Turbulence}, and to the reviews by \citet{Krumholz14PPVI} and \citet{Padoan14PPVI}, for further discussion of this issue.

The second point is that the observed systems show relatively little scatter at SFRs around the average from \citet{Bigiel}. With the exception of a single outlier with particularly low HCN luminosity for its SFR, the majority of the sample of Milky Way objects tends to fall in the range $\eps=0.1\perc-2\perc$ irrespective of our calibration technique. When we calibrate with G, $90\perc$ of the sample falls within this range, and indeed the entire sample save two points falls between the $\eps = 0.1\perc$ and $5\perc$ lines. For calibration without G, $85\perc$ of the sample falls within this range, and more broadly there is a strong constraint between the $\eps = 0.05\perc$ and $5\perc$ lines (albeit with much more scatter). The size of this scatter is consistent with the findings of most other studies that have used different methods to estimate $\eps$ on cloud scales \citep[e.g.][]{Krumholz12a,Christoph13,Evans14a,Salim15,Vutisalchavakul16a,Heyer16a,LeroyM51}, but is substantially smaller than the range reported in \citet{Murray11b} or \citet{Lee16a}. Indeed, the substantial population of objects with $\eps > 10\perc$ reported in \citeauthor{Lee16a} appears to be absent in the massive clump sample. This is significant because one possible explanation for the discrepancy, proposed by \citeauthor{Lee16a}, is that other surveys have focused on smaller star-forming clouds nearby and as a result have missed a class of highly-efficient star-formers at larger distances. The failure of these sources to turn up in the HCN clump samples, which are targeted on massive star-forming regions, casts doubt on this explanation.

On the other hand, unless the factor of few variation in $\eps$ apparent in \autoref{fig:wucomparison} is entirely due to variations in gas temperature or HCN abundances, there is clearly some region-to-region variation in $\eps$. Variations at the factor of few level that we find have in fact been predicted to exist as a result of variations in the Mach numbers, virial parameters, magnetic field strengths, and solenoidal-to-compressive turbulence ratios of molecular clouds \citep[e.g.,][]{Kauffmann13,Schneider13,Christoph13,Christoph16,Jin17,Kainulainen17,Kortgen17}.

\section{Summary and Conclusions}
\label{sec:conclusion}

We post-process a series of high-resolution hydrodynamical simulations of star cluster formation to predict their luminosities in the HCN(1--0) line, and to determine the relationship between HCN luminosity, gas density distribution, and star formation rate. The simulations include a range of physical processes and thus probe a range of modes of star formation, from relatively slow star formation inhibited by strong magnetic fields, turbulence, jets and radiation, to rapid star formation in near free-fall collapse. We find that, nearly independent of the overall star formation rate, HCN emission traces gas with a luminosity-weighted mean density of $0.8-1.7\times 10^4 \cm^{-3}$, and that the conversion between HCN luminosity and mass of gas above $10^4\cm^{-3}$ is $\alphaHCN \approx 14\,\msun/\,(\funnyunits)$. This value is uncertain at the factor of $\sim 2$ level, mainly due to uncertainties in the total HCN abundance. This indeed justifies the perception that HCN(1--0) transitions trace dense gas regions associated with star formation.

We also find that the ratio of star formation rate to HCN emission is strongly correlated with the star formation rate per free-fall time $\eps$, as SFR/$\lhcn \approx 2.0\times10^{-7}\,(\eps/0.01)^{1.1}\,\sfrhcnunits$, with a factor of $\sim3$ systematic uncertainty. Expressed in the more usual terms of the IR--HCN correlation, we find $\lir/\lhcn \approx 1310\,(\eps/0.01)^{1.1}\,\lsun/(\funnyunits)$. Our relation indicates that the observed IR--HCN relation corresponds to a mean star formation rate per free-fall time $\eps \approx 1\perc$, which is highly supportive of typically observed values of $\eps \sim 1\perc$ for similar studies. Of our simulations, only the one with the lowest $\eps$ and the slowest mode of star formation approaches the observed IR--HCN correlation, while those with more rapid modes of star formation all predict far to little HCN luminosity per unit star formation.

We further find that, in a large sample of massive molecular clumps in the Milky Way, the clump-to-clump scatter in $\eps$ is only a factor of a few, with more than $88\perc$ of values falling in the range $\eps = 0.1\perc - 5\perc$ (and this increases to more than $99\perc$ if we calibrate with the G simulation). This result is consistent with findings based on other techniques that $\eps$ varies little from cloud to cloud within the Milky Way. Conversely, we fail to find evidence to support published claims that there is a population of massive star-forming regions with $\eps > 10\perc$.

We conclude that HCN(1--0) transitions are indeed an effective tracer of dense, star-forming gas and that the IR--HCN relation provides a strong constraint on models of star formation that is independent of other methods for determining $\eps$. We suggest that future simulations of star formation check their results against this constraint, and to facilitate such comparisons we provide an implementation of our code to compute HCN luminosities from simulations at \url{http://bitbucket.org/aonus/hcn}.

\section*{Acknowledgements}

M.R.K.~acknowledges funding from the Australian Research Council's Discovery Projects grant DP160100695. C.F.~gratefully acknowledges funding provided by the Australian Research Council's Discovery Projects (grants DP150104329 and DP170100603) and the ANU Futures Scheme, as well as the Australia-Germany Joint Research Cooperation Scheme (UA-DAAD). The simulations presented in this work used high performance computing resources provided by the Leibniz Rechenzentrum and the Gauss Centre for Supercomputing (grants pr32lo, pr48pi and GCS Large-scale project 10391), the Partnership for Advanced Computing in Europe (PRACE grant pr89mu), the Australian National Computational Infrastructure (grant ek9), and the Pawsey Supercomputing Centre with funding from the Australian Government and the Government of Western Australia, in the framework of the National Computational Merit Allocation Scheme and the ANU Allocation Scheme. The simulation software FLASH was in part developed by the DOE-supported Flash Center for Computational Science at the University of Chicago.

\bibliographystyle{mnras}
\bibliography{Bib}

\label{lastpage}
\end{document}